\documentclass[12pt]{article} 
\usepackage{epsfig}
\usepackage{a4}
\usepackage{latexsym}
\usepackage{cite}

\usepackage{pslatex}
\usepackage[latin1]{inputenc}
\usepackage[T1]{fontenc}

\usepackage{color,colordvi}
\def\colour4colour#1{\Blue{#1}}

\newcommand{\gmel}{\raisebox{-0.07cm}{$\stackrel{{\rm M-trf}}{=} $} }
\newcommand{\hspn}{{\hspace{-3mm}}}
\newcommand{\beq}{\begin{equation}}
\newcommand{\eeq}{\end{equation}}
\newcommand{\bea}{\begin{eqnarray}}
\newcommand{\eea}{\end{eqnarray}}
\newcommand{\nn}{\nonumber}
\newcommand{\MSb}{$\overline{\mbox{MS}}$}
\newcommand{\as}{\alpha_{\rm s}}
\newcommand{\ar}{a_{\rm s}}
\newcommand{\ra}{\rightarrow}

\newcommand{\lam}{\lambda}
\newcommand{\GE}{\gamma_{\rm e}}
\newcommand{\GEs}{\gamma_{\rm e}^{\,2}}

\begin{document}
\setlength{\parskip}{0.2cm}
\setlength{\baselineskip}{0.53cm}

\def\Fone{{F_{\:\! 1}}}
\def\Ftwo{{F_{\:\! 2}}}
\def\FL{{F_{\:\! L}}}
\def\F3{{F_{\:\! 3}}}
\def\Qs{{Q^{\, 2}}}
\def\GeV2{{\mbox{GeV}^{\:\!2}}}
\def\DDk{{{\cal D}_{\:\!k}}}
\def\DD#1{{{\cal D}_{\! #1}^{}}}
\def\z#1{{\zeta_{\:\! #1}}}
\def\zss{{\zeta_{2}^{\,2}}}
\def\zst{{\zeta_{3}^{\,2}}}
\def\zts{{\zeta_{2}^{\,3}}}
\def\ca{{C^{}_A}}
\def\cas{{C^{\: 2}_A}}
\def\cat{{C^{\: 3}_A}}
\def\cf{{C^{}_F}}
\def\cfs{{C^{\: 2}_F}}
\def\cft{{C^{\: 3}_F}}
\def\cff{{C^{\: 4}_F}}
\def\nf{{n^{}_{\! f}}}
\def\nfs{{n^{\,2}_{\! f}}}
\def\nft{{n^{\,3}_{\! f}}}
\def\dabc2{{d^{\:\!abc}d_{abc}}}
\def\dabcnc{{{d^{\:\!abc}d_{abc}}\over{n_c}}}
\def\fl11{fl_{11}}
\def\fl02{fl_{02}}
\def\b#1{{{\beta}_{#1}}}
\def\bb#1#2{{{\beta}_{#1}^{\,#2}}}

\def\S(#1){{{S}_{#1}}}
\def\Ss(#1,#2){{{S}_{#1,#2}}}
\def\Sss(#1,#2,#3){{{S}_{#1,#2,#3}}}
\def\Ssss(#1,#2,#3,#4){{{S}_{#1,#2,#3,#4}}}
\def\pqq(#1){p_{\rm{qq}}(#1)}
\def\gfunct#1{{g}_{#1}^{}}
\def\H(#1){{\rm{H}}_{#1}}
\def\Hh(#1,#2){{\rm{H}}_{#1,#2}}
\def\Hhh(#1,#2,#3){{\rm{H}}_{#1,#2,#3}}
\def\Hhhh(#1,#2,#3,#4){{\rm{H}}_{#1,#2,#3,#4}}
\def\Hhhhh(#1,#2,#3,#4,#5){{\rm{H}}_{#1,#2,#3,#4,#5}}

\def\gqqz{\gamma_{\,\rm qq}^{\,(0)}}
\def\gnso{\gamma_{\,\rm v}^{\,(1)}}
\def\gnst{\gamma_{\,\rm v}^{\,(2)}}
\def\ctqo{c_{3,\pm}^{(1)}}
\def\ctnt{c_{3,-}^{(2)}}
\def\ctnd{c_{3,-}^{(3)}}
\def\atnt{a_{3,-}^{(2)}}
\def\btqo{b_{3,\pm}^{(1)}}
\def\atqo{a_{3,\pm}^{(1)}}

\begin{titlepage}
\noindent
DESY 09-022 \hfill {\tt arXiv:0902.2342 [hep-ph]}\\
SFB/CPP-09-16 \\
LTH 821 \\[1mm]
February 2009 \\
\vspace{1.5cm}
\begin{center}
\Large
{\bf Threshold Resummation of the Structure Function $F_L$}\\
\vspace{2.5cm}
\large
S. Moch$^{\, a}$ and A. Vogt$^{\, b}$\\
\vspace{1.5cm}
\normalsize
{\it $^a$Deutsches Elektronensynchrotron DESY \\
\vspace{0.1cm}
Platanenallee 6, D--15738 Zeuthen, Germany}\\
\vspace{0.5cm}
{\it $^c$Department of Mathematical Sciences, University of Liverpool \\
\vspace{0.1cm}
Liverpool L69 3BX, United Kingdom}\\[2.5cm]
\vfill
\large
{\bf Abstract}
\vspace{-0.2cm}
\end{center}
The behaviour of the quark coefficient function for the longitudinal structure
function $\FL$ in deep-inelastic scattering is investigated for large values of 
the Bjorken variable $x$. We combine a highly plausible conjecture on the 
large-$x$ limit of the physical evolution kernel for this quantity with our 
explicit three-loop results to derive the coefficients of the three leading 
large-$x$ logarithms, $\as^{\,n}\: \ln^{\,2n-1-k} (1-x)$, $k\,=\, 1,\: 2,\: 3$,
to all orders in the strong coupling constant $\alpha_s$. Corresponding results
are derived for the non-$\cf$ part of the gluon coefficient function suppressed
by a factor $\,1-x$, and for the analogous subleading $(1-x)\,\ln^{\,k}(1-x)$
contributions in the quark case.
Our results appear to indicate an obstacle for an exponentiation with a higher 
logarithmic accuracy.
   
\vfill
\end{titlepage}
%
% -----------------------------------------------------------------------------
%
\noindent
%Structure functions in inclusive deep-inelastic scattering (DIS) provide the 
%most important (and accessible) laboratory for studying higher-order 
%
Structure functions in deep-inelastic scattering (DIS) provide an important (and
accessible, via forward Compton amplitudes) laboratory for studying higher-order
effects in perturbative QCD. Indeed, they are presently the only observables 
depending on a dimensionless variable (Bjorken-$x$ in the case at hand) 
for which Feynman diagram calculations have been extended to the third order in 
the strong coupling $\as$ 
\cite{Moch:2004pa,Vogt:2004mw,Moch:2004xu,Vermaseren:2005qc,Moch:2008xx}. 
Such calculations are not only relevant phenomenologically, but also open up 
ways to new results for different quantities. For instance, a direct line runs 
from an observation on subleading large-$x$ logarithms at three loop 
\cite{Moch:2004pa,Vogt:2004mw} via its interpretation in Ref.~\cite
{Dokshitzer:2005bf} to first results on the third-order splitting functions for 
the final-state parton fragmentation \cite{Mitov:2006ic,Moch:2007tx}.

In the present letter we study the same class of large-$x$ contributions, 
$\as^{\,n}\, \ln^{\,k}(1-x)$, to the higher-order quark coefficient functions 
\cite{SanchezGuillen:1991iq,vanNeerven:1991nn,Moch:1999eb,Moch:2004xu,%
Vermaseren:2005qc} for the longitudinal structure function $\FL$ (an analogous 
investigation of $\Ftwo$ and $\F3$ will be presented elsewhere 
\cite{MV4}$\:\!$) where these logarithms form the leading terms at $x \ra 1$.
Such contributions have been addressed before in Refs.~\cite
{Akhoury:1998gs,Akhoury:2003fw,Grunberg:2007nc,Laenen:2008gt}, but no explicit 
all-order predictions have been presented so far for any coefficient function 
beyond the leading logarithms. 
This situation for the leading large-$x$ behaviour of $\FL$ is in striking 
contrast to that for $\Ftwo$ and $\F3$ where the soft-gluon exponentiation 
\cite{Sterman:1987aj,Catani:1989ne} 
is known to the next-to-next-to-next-to-leading logarithmic accuracy and 
predicts the leading seven term to all orders in $\as$ \cite{Moch:2005ba}.

The (flavour non-singlet) quark coefficient functions $C_{a,\rm ns}$ provide 
the connection between the structure functions $F_{a,\,\rm ns}$ and the 
corresponding quark distributions $q_{\rm ns\,}^{}$, 
\bea
\label{Fns-cq}
  {\cal F}_{a=2,L}(x,Q^2) &\! \equiv\! &
   x^{\,-1} F_{a,\,\rm ns}(x,Q^2) \:\: =\:\: C_{a,\rm ns}(x,\as) \,\otimes\, 
   q_{\rm ns}^{}(x,Q^2)
 \nn \\[1mm] &\! =\! & \Big[
  (1-\delta_{aL}) \delta(1-x) + \sum_{n=1}\:\ar^{\,n}\: c^{\,(n)}_{\,a,q}(x)
  \Big] \,\otimes\, q_{\rm ns}(x,Q^2) \:\: ,
\eea
where $\otimes$ stands for the standard Mellin convolution. The renormalization 
and factorization scales $\mu_{\:\!\rm r\,}$ and $\mu_{\:\!\rm f\,}^{}$ have 
been set to the physical hard scale $\Qs$ in Eq.~(\ref{Fns-cq}), and the 
expansion parameter is normalized as $\ar = \as/(4\pi)$. 
The large-$x$ expansion of the \MSb\ coefficient function for $\FL$ reads
\bea
\label{CLexp}
  C_{\,L,\rm ns}(\as, x) \!\!& = & 
     \sum_{n=1}\:\ar^{\,n}\: c^{\,(n)}_{\,L,q}(x) 
 \nn \\[-1mm]  & = &
     \sum_{n=1}\:\ar^{\,n}\: \Bigg\{ \,
%         \sum_{k=0}^{2n-2} \bar{c}^{\,(n)}_{\,L,\,k}\; \ln^{\,k}(1-x)
%         \: + \: O\Big( \,(1-x) \ln^{\,2n-2}(1-x) \Big) \Bigg\}
     \sum_{k=0}^{2n-2} \ln^{\,k}(1-x) \Big[ \,\bar{c}^{\,(n)}_{\,L,\,k} \:+ \:  
          (1-x)\, \bar{d}^{\,(n)}_{\,L,\,k} \: + \: O\left( (1-x)^2\, \right) 
          \Big] \Bigg\}
 \quad \nn \\ & \!\gmel\! & {1 \over N}\,
     \sum_{n=1}\:\ar^{\,n}\: \Bigg\{ \,
%         \sum_{k=0}^{2n-2} c^{\,(n)}_{\,L,\,k}\; \ln^{\,k}N
%         \: + \: O\Big( \, {1 \over N}\: \ln^{\,2n-2} N \Big)\Bigg\}
          \sum_{k=0}^{2n-2} \ln^{\,k}N \Big[ \, c^{\,(n)}_{\,L,\,k} \:+ \:
          {1 \over N}\, d^{\,(n)}_{\,L,\,k} \: + \: 
          O\left( {1 \over N^{\,2}} \right) \Big] \Bigg\}
 \:\: . 
\eea
Here and below $\;\gmel\; $ indicates that the right-hand-side is the Mellin
transform of the previous expression. 
The leading $x$- and $N$-space coefficients $\bar{c}^{\,(n)}_{\,L,\,k}$ and 
$c^{\,(n)}_{\,L,\,k}$ in Eq.~(\ref{CLexp}) are related via
\bea
\label{Mtrf}
  (-1)^k \int_0^1 dx\, x^{\,N-1} \ln^{\,k}(1-x) &\! =\! &
    k!\: {1 \over N}\: S_{\underbrace{1,\dots,1}_{\scriptstyle k}}(N) 
 \nn \\[-1mm]  &\! =\! &
    {1 \over N}\, \ln^{\,k} \widetilde{\!N} \: +\: \sum_{l=2}^{k} 
    \frac{k!}{l (k-l)!}\: \bar{\zeta}_{\,l}\, {1 \over N}\: \ln^{\,k-l} 
    \widetilde{\!N} \: +\: O\Big( \, {1 \over N^{\,2}}\, \ln^{\,k-1} N \Big) 
 \quad
\eea
with $\:\widetilde{\!N} \,=\, N e^{\,\GE}$ and the Riemann-zeta combinations 
$\bar{\zeta}_{2,3} \,=\, \zeta_{2,3}$, $\,\bar{\zeta}_{4} \,=\, \zeta_{4} + 
{1 \over 2} \zeta_2^2$, $\,\bar{\zeta}_{5} \,=\, \zeta_{5} + {5 \over 6} 
\zeta_2 \zeta_3$ etc. See Refs.~\cite{Vermaseren:1998uu,Blumlein:1998if} for 
the notation and properties of the harmonic sums $S_{m_1,\dots,\,m_k}(N)$. 

It is convenient, both phenomenologically (especially for $\Ftwo$ and $\F3$)
and theoretically, to express the scaling violations of non-singlet structure 
functions in terms of these structure functions themselves. This explicitly 
eliminates any dependence on the factorization scheme and the scale
$\mu_{\:\!\rm f\,}^{}$. The corresponding `physical evolution kernels' $K_a$ can 
be derived for $\mu_{\:\!\rm r}^{\,2} = Q^{\,2}$ by differentiating 
Eq.~(\ref{Fns-cq}) with respect to $Q^{\,2}$ by means of the evolution equations 
for $\,\ar = \as/(4\pi)$ and~$q_{\rm ns}^{}$, 
\bea
\label{arun}
  \frac{d\,\ar}{d \ln \Qs} &\! = \!& \beta(\ar) \:\: = \:\:
  - \beta_0\, \ar^{\,2} - \beta_1\, \ar^{\,3} - \:\ldots 
  \:\: , \qquad \beta_0 \: = \: \frac{11}{3}\:\ca - \frac{2}{3}\:\nf \:\: ,
 \\
\label{qrun}
 \frac{d\,q_{\rm ns}^{}}{d \ln \Qs} &\! = \!& P_{\rm ns} \otimes q_{\rm ns}^{}
 \:\: = \:\: 
 \sum_{n=1} \ar^{\,n}\, A_n [1-x]_+^{-1} \otimes q_{\rm ns}^{} + \ldots  
 \;\;\gmel\;\; - \sum_{n=1} \ar^{\,n}\, A_n \ln N + \:\ldots \;\; . \quad
\eea
The `cusp anomalous dimension' 
$A(\ar) \,=\, A_1\,\ar + A_2\,\ar^{\,2} + \ldots\:$ with $A_1 \,=\, 4\,\cf\,$ 
has been calculated to order $\as^{\,3}$ \cite{Moch:2004pa}. Finally using the 
inverse of Eq.~(\ref{Fns-cq}) to eliminate $q_{\rm ns}^{}$ leads to the 
evolution equations
\bea
\label{Fevol}
  \frac{d}{d \ln Q^{\,2}} \; {\cal F}_a
  &\!\!\! =\!\! &
  \bigg\{ P_{\rm ns}(\ar) + \beta(\ar)\: \frac{d}{d \ar} \ln\, C_a(\ar) \!
  \bigg\} \otimes \: {\cal F}_a
%\nn\\[1.5mm]
%  & \!\! \equiv \! & \;
  \:\: = \:\:
  K_a \otimes\, {\cal F}_a \:\: \equiv \:\:
    \sum_{n=1} \, \ar^{\, n}\, K_a^{\,(n)} \otimes \, {\cal F}_a
 \:\: .
\eea
Inserting the coefficients known from 
Refs.~\cite{Moch:2004xu,Vermaseren:2005qc},
the same leading-logarithmic behaviour for both $\Ftwo$ and $\FL$, viz
\bea
\label{Kxto1}
  K_a^{\,(n)}(x) &\! =\! & \mbox{}
    A_1 (-\beta_0)^{\,n-1} \left[ \,\frac{\ln^{\,n-1}(1-x)}{1-x} \right]_+ 
  + \; O\left( \left[ \,\frac{\ln^{\,n-2}(1-x)}{1-x} \right]_+ \right)
 \nn\\[-0.5mm] & \!\gmel\! &
%  \;\gmel\;
   - \, {A_1 \beta_0^{\,n-1} \over n} \; \ln^{\,n} N
  \; + \; O \left( \ln^{\,n-1} N \:\!\right) \:\: ,
\eea
is established to $\,n=4\,$ for $\Ftwo$ and $\,n=3\,$ for $\FL$. For $\,\Ftwo$
the soft-gluon resummation~\cite{Sterman:1987aj,Catani:1989ne,Moch:2005ba},
\beq
\label{C2res}
  C_{\,2,\rm ns}(N,\ar) \:\: = \:\: g_2^{(0)}(\ar)\, 
  \exp \left[ L g_2^{(1)}(\ar\,L) + g_2^{(2)}(\ar\,L) + \dots \right] 
\:\: , \quad
  g_2^{(i)}(\lam) \:\: = \:\: \sum_j g_{2j}^{(i)} \:\lam^{j}
\:\: ,
\eeq
($L \,\equiv\, \ln\, N$)
guarantees Eq.~(\ref{Kxto1}) to all orders in $\as$ \cite{vanNeerven:2001pe}. 
It is crucial that the physical kernel, unlike the coefficient functions, 
receives only this single-logarithmic higher-order enhancement for $x \ra 1\,$.

We are now, finally, in a position to state the conjecture announced in the
abstract.
It is {\bf (a)} that this single-logarithmic enhancement remains true for $\FL$
beyond order $\as^{\,3}$ and {\bf (b)} that Eq.~(\ref{Kxto1}) holds to (at 
least) $n=4$ also for $\FL$. {\bf (a)} implies that that there is an 
exponentiation as Eq.~(\ref{C2res}) (but, of course, with an overall prefactor 
$N^{-1}$) also for $\FL$ with some functions $g_L^{(i)}$. {\bf (b)} 
additionally requires that the leading logarithmic functions $g_a^{(1)}$ are 
actually the same for $a=2$ and $a=L$ to (at least) order $\as^{\,3}$. We 
consider the results of Refs.~\cite
{Akhoury:1998gs,Akhoury:2003fw,Laenen:2008gt} as sufficient evidence for these
natural assumptions generalizing our fixed-order results.
In particular, it may be expected that the new approach of 
Ref.~\cite{Laenen:2008gt} will facilitate a full proof in the future.

Inserting Eqs.~(\ref{CLexp}), (\ref{arun}) and (\ref{qrun}) into 
Eq.~(\ref{Fevol}) and imposing the vanishing of the resulting $\as^{\,n} 
\ln^{\,2n-2}$ and $\as^{\,n} \ln^{\,2n-3}$ contributions to $K_L^{\,(n)}$ at 
$n \geq 4$ fixes the coefficients of the two highest logarithms in 
Eq.~(\ref{CLexp}) to all orders $n$ in $\as\,$ 
(with $\theta_{nj} = 1$ for $n\geq j$ and $\theta_{nj} = 0$ else)$\,$: 
\bea
\label{cL-LL}
  c^{\,(n)}_{L,\,2n-2} &\! =\! & 
    2(2\cf)^n \: \frac{1}{(n-1)!} \:\: ,
 \\
\label{cL-NhLL}
  c^{\,(n)}_{L,\,2n-3} &\! =\! &
    c^{\,(2)}_{\,L,\,1}\, (2\cf)^{n-2}\, \frac{\theta_{n2}}{(n-2)!} \: + \:
    {2\beta_0 \over 3}\,  (2\cf)^{n-1}\, \frac{\theta_{n3}}{(n-3)!} \:\: .
% \\[1mm]
\eea
We have conjectured Eq.~(\ref{cL-LL}) before \cite{Moch:2004xu} on the basis
of the explicit calculations for $n \leq 3$ and the results of Refs.~\cite
{Akhoury:1998gs,Akhoury:2003fw}. To the best of our knowledge, Eq.~(\ref
{cL-NhLL}) has not been written down before. Furthermore the vanishing of the 
$\as^{\,n} \ln^{\,2n-4}$ contributions to $K_L^{\,(n)}$ at $n \geq 5$ yields
\bea
\label{cL-NLL}
  c^{\,(n)}_{L,\,2n-4} &\! =\! &
    c^{\,(3)}_{\,L,\,2}\, (2\cf)^{n-3}\, \frac{\theta_{n3}}{(n-3)!} \: + \:
    {\beta_0 \over 3}\, c^{\,(2)}_{\,L,\,1}\, (2\cf)^{n-3}\, 
       \frac{\theta_{n4}}{(n-4)!} \: - \:
  c^{\,(2)}_{L,\,0}\, (2\cf)^{n-2}\, \frac{(n-3)\theta_{n4}}{(n-2)!}
\quad \nn \\[1mm] &  & \mbox{} + \:
   {\beta_0^2 \over 9}\, (2\cf)^{n-2} \frac{\theta_{n5}}{(n-5)!}  \: - \:
   {2 \over 3\beta_0}\, K_L^{\,(4)} \Big|_{\ln^{\,4}N}\, (2\cf)^{n-3}\, 
       \frac{\theta_{n4}}{(n-4)!} \:\: .
\eea
The last line includes the leading term of the physical kernel at order 
$\as^{\,4}$, i.e., we have not included conjecture {\bf (b)} in the derivation
of Eq.~(\ref{cL-NLL}). 
After inserting Eq.~(\ref{Kxto1}) for $a=L$ and $n=4$, i.e., applying also 
{\bf (b)}, we arrive at a definite prediction also for the third tower 
(\ref{cL-NLL}) of logarithms, thus reaching the predictive power of a 
next-to-leading logarithmic exponentiation, cf.~Ref.~\cite{Moch:2005ba}.
The other coefficients in Eqs.~(\ref{cL-LL}) -- (\ref{cL-NLL}) can be extracted
from the loop calculations in 
Refs.~\cite{SanchezGuillen:1991iq,vanNeerven:1991nn,Moch:1999eb,Moch:2004xu,% 
Vermaseren:2005qc}, 
\bea
\label{cL10}
  c^{\,(1)}_{L,\,0} &\! =\! &
       4\, \cf
 \\[0.5mm]
\label{cL21}
  c^{\,(2)}_{L,\,1} &\! =\! &
     \cf \ca \Bigg[ \, 
            {92 \over 3} 
          - 16\, \z2 \Bigg] 
     \: - \:  \cfs \, [ 
            36 - 32\, \z2 - 16\, \GE 
            ] 
     \: - \: {8 \over 3}\: \cf \nf 
 \\[1mm]
\label{cL20}
 c^{\,(2)}_{L,\,0} &\! =\! & \mbox{}
        - \: \cfs \, \left[ \,
            34
          + 40\, \z2
          - 48\, \z3
          + 36\, \GE
          - 32\, \GE \z2
          - 8\,  \GEs
          \,\right]
\nn \\[0.5mm] & & \mbox{}
        + \: \cf \ca \Bigg[ \,
            {430 \over 9} 
          + 16\, \z2 
          - 24\, \z3 
          + {92 \over 3}\: \GE 
          - 16\,\GE \z2
          \Bigg]
     \: - \: \cf \nf \, \Bigg[ \,
            {76 \over 9} 
          + {8 \over 3}\: \GE
          \Bigg]
 \\[1mm]
\label{cL32}
 c^{\,(3)}_{L,\,2} &\! =\! & \mbox{}
        - \: \cft \, \left[ \,
            34 
          - 16\, \z2 
          + 32\, \z3 
          + 216\, \GE 
          - 192\, \GE \z2 
          - 48\, \GEs
          \,\right]
     \: + \: {16 \over 9}\: \cf \nfs 
\nn \\ & & \mbox{}
        - \: \cfs \ca \Bigg[ \,
            {530 \over 9} 
          - 80\, \z2 
          - 80\, \z3 
          - {640 \over 3}\: \GE  
          + 96\, \GE \z2
          \Bigg]
     \: - \: \cf \ca \nf \Bigg[ \,
            {320 \over 9} 
          - 16\, \z2 
          \Bigg]
\quad \nn \\[1mm] & & \mbox{}
        + \: \cf \cas \Bigg[ \,
            {1276 \over 9} 
          - 56\, \z2
          - 32\, \z3 
          \Bigg]
     \: + \: \cfs \nf \Bigg[ \,
            {92 \over 9} 
          - 32\, \z2 
          - {64 \over 3}\, \GE
          \Bigg]
 \;\; .
\eea
 
Inserting Eqs.~(\ref{cL10}) -- (\ref{cL32}) into Eqs.~(\ref{cL-LL}) --
(\ref{cL-NLL}) and transforming back to $x$-space, one arrives at the four-loop
prediction (using $L_x \,\equiv\, \ln (1-x)$ for brevity)
\bea
\label{cL4x}
  c^{\,(4)}_{L,q}(x) &\!\! =\! &
    {16 \over 3}\: \cff\: L_x^{\,6}
  \: + \: \Bigg\{ 
    [\, 72 - 64\, \z2 ] \,\cff 
    - \Bigg[ {728 \over 9} - 32\, \z2 \Bigg] \: \cft \ca 
    + {80 \over 9}\, \cft \nf 
  \Bigg\} \: L_x^{\,5}
\nn \\[1mm] & & \mbox{\hspn\hspn}
  + \: \Bigg\{ [\, 32\, \z2 - 160\, \z3 ] \,\cff
    - \Bigg[ {904 \over 3} - {1856 \over 9}\: \z2 - 208\, \z3 \Bigg]\: \cft \ca
    + \Bigg[ {160 \over 3} - {704 \over 9}\: \z2 \Bigg]\: \cft \nf
\nn \\[1mm] & & \mbox{\hspn} \;
    + \Bigg[ {3388 \over 9} - {1360 \over 9}\: \z2 - 64\,\z3 \Bigg]\: \cfs \cas
    - \Bigg[ {880 \over 9} - {352 \over 9}\: \z2 \Bigg]\: \cfs \ca \nf
    + {16 \over 3}\: \cfs \nfs
  \Bigg\} \: L_x^{\,4} \quad
\nn \\[2mm] & & \mbox{\hspn\hspn} + \; O(L_x^{\,3}) \;\; .
\eea
This result will become useful also outside the large-$x$ region in combination
with a future generalization of Ref.~\cite{Larin:1997wd} to low fixed-$N$ 
moments at order $\as^{\,4}$, since fewer moments will be needed for a useful
$x$-space approximation analogous to Ref.~\cite{vanNeerven:2001pe}.

For future applications and possible extensions to next-to-next-to-leading
logarithmic accuracy, it is useful
to reformulate Eqs.~(\ref{cL10}) -- (\ref{cL32}) in terms of the exponentiation
coefficients $g_{ij} \,\equiv\, g_{Lj}^{(i)}$. For this purpose we adapt 
Eq.~(14) of Ref.~\cite {Vogt:1999xa} to the present case with 
$\,g_L^{(0)} = N^{-1} [\,\ar c^{\,(1)}_{L,\,0} \,+\, O(\ar^{\,2})]$ instead 
of $\,g_2^{(0)} = 1 + O(\ar)$, yielding
\bea
\label{tower1}
  c^{\,(n)}_{L,\,2n-2\,}/(4\,\cf) &\! =\! & 
    \frac{g_{11}^{\,n-1}}{(n-1)!} 
 \:\: , \\[1mm]
\label{tower2}
  c^{\,(n)}_{L,\,2n-3\,}/(4\,\cf) &\! =\! &
    \frac{\theta_{n2}\, g_{11}^{\,n-2}}{(n-2)!}\, g_{21}^{}
    \: + \;\frac{\theta_{n3}\, g_{11}^{\,n-3}}{(n-3)!}\, g_{12}^{}
 \:\: , \\[1mm]
\label{tower3}
  c^{\,(n)}_{L,\,2n-4\,}/(4\,\cf) &\! =\! &
      \frac{\theta_{n2}\, g_{11}^{\,n-2}}{(n-2)!}\, g_{01}^{}
    \: + \;\frac{\theta_{n3}\, g_{11}^{\,n-3}}{(n-3)!}\, \Big( g_{22}^{}
      + \frac{1}{2} g_{21}^{\,2} \Big)
    \: + \;\frac{\theta_{n4}\, g_{11}^{\,n-4}}{(n-4)!}\, \Big( g_{13}^{}
      + g_{12}^{} g_{21}^{} \Big)
 \quad \nn \\ & & \mbox{\hspn}
    + \frac{\theta_{n5}\, g_{11}^{\,n-5}}{2(n-5)!}\, g_{12}^{\,2}
 \;\; . 
\eea
Eqs.~(\ref{cL-LL}) and (\ref{cL-NhLL}) are obviously compatible with 
Eqs.~(\ref{tower1}) and (\ref{tower2}). Also Eq.~(\ref{cL-NLL}) can be recast
in the form (\ref{tower3}) by suitably combining the first and the last term in
the first line. The comparison of the two sets of expressions then leads to
\bea
\label{g1k}
  g_{11}^{} &\! =\! & 2\,\cf 
\:\: , \quad 
  g_{12}^{} \:\: = \:\: \frac{2}{3}\: \beta_0\, \cf
\:\: , \quad 
  g_{13}^{} \:\: = \:\: \frac{1}{3}\: \beta_0^{\,2}\, \cf 
\:\: , \\[1mm]
\label{g21}
  g_{21}^{} &\! =\! & \beta_0 \:+\: 4\,\GE\cf \:-\: \cf
    \:+\: (4 - 4\,\zeta_2) (\ca - 2\cf)  
\eea
and
\bea
\label{g22}
  g_{22}^{} &\! =\! &
        - 32\, \cfs \, \left[ \,
            1 
          - 3\, \z2
          + \z3
          + \zss
          \,\right]
     \: + \: \cf \ca \, \Bigg[ \,
          {547 \over 18}
          - {256 \over 3}\: \z2
          + 32\, \z3
          + 32\, \zss
          + {22 \over 3}\, \GE
          \Bigg]
     \: + \: {2 \over 9}\: \nfs 
\nn \\[0.5mm] & & \mbox{}
        + \: \cas \Bigg[ \,
            {109 \over 18}
          + {50 \over 3}\: \z2
          - 8\, \z3
          - 8\, \zss
          \Bigg]
     \: + \: \cf \nf \, \Bigg[ \,
            {7 \over 9}
          - {8 \over 3}\: \z2
          - {4 \over 3}\, \GE
          \Bigg]
        - \: \ca \nf \Bigg[ \,
            {34 \over 9}
          - {4 \over 3}\: \z2
          \Bigg]
%\nn \\[1mm] & & \mbox{}
%        - \: \ca \nf \Bigg[ \,
%            {34 \over 9}
%          - {4 \over 3}\: \z2
%          \Bigg]
%     \: + \: {4 \over 9}\: \nfs 
\nn \\[1.5mm] &\! =\! & 
         {1 \over 2}\, (\, \beta_0 \,g_{21}^{} + A_2 ) 
       \: - \: 8\, (\ca - 2\cf)^2 \Big( 1 - 3\,\z2 + \z3 + \zss\, \Big)
\eea 
with the two-loop cusp anomalous dimension $\,A_2 = 8\,C_F K$,~ 
$K = ( 67/18 - \z2 )\, C_A - 5\,\nf/9$ \cite{Kodaira:1981nh}.

Some comments are in order here: As expected from the discussion below 
Eq.~(\ref{C2res}) the relations (\ref{g1k}), with the value of $g_{13}^{}$ due 
to conjecture {\bf (b)}, are identical to Eq.~(9) in Ref.~\cite{Vogt:1999xa}. 
The third and first term of Eq.~(\ref{g21}) are identical, up to a trivial
normalization factor, to $\gamma^{}_{J^{\,\prime}}$ in Eq.~(16) of 
Ref.~\cite{Akhoury:1998gs} --- see also Eq.~(48) of Ref.~\cite{Akhoury:2003fw} 
and note that the presence of $\GE$ in Eq.~(\ref{g21}) results from our use of 
$L \,\equiv\,\ln\, N$ instead of $\ln\, \widetilde{\!N}$ in Eq.~(\ref{C2res})
(keeping $\GE$ facilitates some easy checks).

As shown by the last line of Eq.~(\ref{g22}), the coefficient $g_{22}^{} 
\,\equiv\, g_{L2}^{(2)}$ in the expansion of $g_{L}^{(2)}(\ar L)$ does not 
follow the pattern of the resummation for $\Ftwo$ which would demand 
$g_{22}^{} = 1/2\: (\:\!\beta_0 \,g_{21}^{} + A_2)$ (cf., e.g., Eq.~(10) of 
Ref.~\cite{Vogt:1999xa}$\,$), i.e., the absence of the `non-planar' 
$(\ca - 2\,\cf)^2$ part in Eq.~(\ref{g22}). Hence also $g_{L3}^{(2)}$ cannot be
predicted at this point (if at all -- consider the $\zeta_3$ contributions to 
Eqs.~(\ref{cL20}), (\ref{cL32}) and (\ref{g22})$\,$) from lower-order 
information. Consequently $c^{\,(3)}_{L,\,1}$, known from Ref.~\cite
{Vermaseren:2005qc}, can be used to derive $g_{L1}^{(3)}$, but not (yet) the 
fourth tower $c^{\,(n)}_{L,\,2n-4\,}$ of logarithms at orders $n \geq 4$.

Let us briefly turn to the gluon coefficient function $C_{\,L,g}$ for the
structure function $\FL$ which is suppressed by an(other) order in $(1-x)$. From the
third-order results in Refs.~\cite{Moch:2004xu,Vermaseren:2005qc} we extract
that
\beq
\label{CLg-LL}
  C_{\,L,g}(\as, x) \:\: = \:\: \sum_{n=1}\:\ar^{\,n} 
  \Bigg\{ 8\,\nf\: \frac{(2\,\ca)^{n-1}}{(n-1)!}\;  
     {1 \over N^{\,2}}\: \ln^{\,2n-2} N 
  \: + \: O\bigg( {1 \over N^{\,2}}\: \ln^{\,2n-3} N \bigg) \Bigg\}
\eeq
holds for the first three terms of the expansion in powers of $\as$ 
(for $n=1$ one obviously has $O(N^{\,-3})$ instead of the last term in 
Eq.(\ref{CLg-LL})$\,$). 
The generalization to all $n$ can be obtained via the physical kernel for the 
`non-singlet' (no gluons emitted from quarks, i.e., only the $C_A^{\,k} 
\, n^{\,n-k}_{\! f}$ terms are kept) gluon contribution to $\FL$ (cf.~also 
Ref.~\cite{Moch:2007tx}$\,$).
In fact, in this unphysical limit our whole previous treatment of the 
(non-singlet -- the pure-singlet part does not contribute at the present
accuracy) quark coefficient function can be carried over to the gluon case in 
an obvious manner. Since they might be of theoretical interest at some point, 
we present here the relations corresponding to Eqs.~(\ref{g1k}) -- (\ref{g22}):
\bea
\label{gg1k}
  g_{11,g}^{} &\! =\! & 2\,\ca
\:\: , \qquad
  g_{12,g}^{} \:\: = \:\: \frac{2}{3}\: \beta_0\, \ca
\:\: , \qquad
  g_{13,g}^{} \:\: = \:\: \frac{1}{3}\: \beta_0^{\,2} \ca
\:\: , \\[0.5mm]
\label{gg21}
   g_{21,g}^{} &\! =\! & (8 + 4\,\GE) \,\ca
%\:\: , \\[1mm]
%\label{gg22}
%   g_{22,g}^{} &\! =\! & 
 \:\: , \quad  g_{22,g}^{} \:\: = \:\: 
   {1 \over 2}\, [\, \beta_0 \,g_{21,g}^{} + A_{2,g} ] 
   \:+\; \cas\, ( 18 - 8\,\z2 ) 
\eea
with $\, A_{2,g} \,=\, \ca/\cf\; A_2$. The situation for $\,g_{22,g}^{}$ is 
analogous, if simpler in terms of the $\zeta$-function, to that in 
Eq.~(\ref{g22}) discussed above.

Returning to the quark case, we note that also the subleading $\ln^{\,k}(1-x)$
or $\,N^{\,-1}\ln^{\,k}N$ contributions to the physical kernel (\ref{Fevol})
show only a single logarithmic higher-order enhancement, again in contrast to 
the corresponding $(1-x)\,\ln^{\,k}(1-x)$ or $\,N^{\,-2}\ln^{\,k}N$ terms in 
the coefficient function. I.e.,
\beq
\label{Kxto1s}
  K_a^{\,(n)}(x)\Big|_{\ln^{\,k}(1-x)} 
  \:\:\: \gmel \;\;\; K_a^{\,(n)}(N)\Big|_{N^{-1}\ln^{\,k}N}
  \:\: = \:\: 0 \quad \mbox{ for }  \quad k \:\geq\: n 
\eeq
where, as before, $n$ stands for the order in $\as$. Also Eq.~(\ref{Kxto1s}) 
is the result of the fixed-order calculations 
\cite{Moch:2004xu,Vermaseren:2005qc,Moch:2008xx} at $\,n \leq 4\,$ for 
$\,a=2,\:3$ -- the missing four-loop splitting function does not contribute at
this logarithmic level \cite{Dokshitzer:2005bf} --
and at $\,n \leq 3\,$ for $\,a=L$. It appears almost obvious that 
also this result holds to all orders. Hence we can predict, completely 
analogous to \mbox{Eqs.~(\ref{cL-LL}) -- (\ref{cL-NLL})}, the three sub-leading
coefficients $d^{\,(n)}_{L,\,2n-1-k}$, $k = 1,\: 2,\:3$ in Eq.~(\ref{CLexp}) 
at all higher orders, and `postdict' $d^{\,(2)}_{L,2}$ (and $d^{\,(3)}_{L,4}$ 
and $d^{\,(3)}_{L,3}\,$) from first- (and second-)$\,$order coefficients.
Due to $d^{\,(1)}_{L,\,0} \,=\, - c^{\,(1)}_{L,\,0}$ the overall signs are
opposite to those in Eqs.~(\ref{cL-LL}) -- (\ref{cL-NLL}), and the most compact
representation of the results is obtained via the sum of the corresponding 
$\ln^{\,k}N$ and $N^{\,-1} \ln^{\,k}N$ coefficients. It reads
\bea
\label{dL-LL}
  d^{\,(n)}_{L,\,2n-2} &\! =\! & \mbox{} - c^{\,(n)}_{L,\,2n-2}
 \:\: ,
 \\[1mm]
\label{dL-NhLL}
  d^{\,(n)}_{L,\,2n-3} &\! =\! & \mbox{} - c^{\,(n)}_{L,\,2n-3} \: + \:
    \Big\{ d^{\,(2)}_{\,L,\,1} +  c^{\,(2)}_{\,L,\,1} \Big\}
    \, (2\cf)^{n-2}\, \frac{\theta_{n2}}{(n-2)!} \:\: ,
\\[1mm]
\label{dL-NLL}
  d^{\,(n)}_{L,\,2n-4} &\! =\! & \mbox{} - c^{\,(n)}_{L,\,2n-4} \: + \:
    \Big\{  d^{\,(3)}_{\,L,\,2} + c^{\,(3)}_{\,L,\,2} \Big\} 
    \, (2\cf)^{n-3}\, \frac{\theta_{n3}}{(n-3)!} 
\\ &  & \mbox{\hspn} + \:
    \Big\{ d^{\,(2)}_{\,L,\,1} + c^{\,(2)}_{\,L,\,1} \Big\} 
    {\beta_0 \over 3}\, (2\cf)^{n-3}\, \frac{\theta_{n4}}{(n-4)!} \: - \:
    \Big\{ d^{\,(2)}_{L,\,0} -  c^{\,(2)}_{L,\,0} \Big\} 
    \, (2\cf)^{n-2}\, \frac{(n-3)\theta_{n4}}{(n-2)!}
\nn \:\: . \quad
\eea
The lower-order coefficients entering these relations read 
\bea
%\label{dL10}
%  d^{\,(1)}_{L,\,0} &\! =\! & \mbox{} - c^{\,(1)}_{L,\,0} 
% \\[0.5mm]
\label{dL21}
  d^{\,(2)}_{L,\,1} &\! =\! & \mbox{} - c^{\,(2)}_{L,\,1} 
     \: - \: 8\,\cfs \:\: ,
 \\[1mm]
\label{dL20}
  d^{\,(2)}_{L,\,0} &\! =\! & \mbox{} - c^{\,(2)}_{L,\,0}
     \: + \: \cfs \, \left[ \,
            14
          + 16\, \z2
          - 8\,  \GE
          \,\right] 
     \: - \: \cf \ca \left[ \,
            14
          + 8\,\z2
          \,\right]
     \: + \: 4\, \cf \nf  
\:\: , \\[1mm]
\label{dL32}
 d^{\,(3)}_{L,\,2} &\! =\! & \mbox{} - c^{\,(3)}_{L,\,2}
     \: + \: \cft \, \left[ \,
            148 
          - 32\, \z2 
          - 48\, \GE
          \,\right]
        - \: \cfs \ca \left[ \,
            104 
          - 16\, \z2 
          \right]
     \: + \: 16\, \cfs \nf 
\:\: .
\eea
 
Obviously it is possible to recast also Eqs.~(\ref{dL-LL}) -- (\ref{dL-NLL})
into an exponential form analogous to Eqs.~(\ref{tower1}) -- (\ref{g22}).
The corresponding leading-logarithmic function $\tilde{g}_1^{}$ is the same as 
in Eqs.~(\ref{g1k}), while $\tilde{g}_{21}^{}$ and $\tilde{g}_{22}^{}$ differ 
from their counterparts in Eqs.~(\ref{g21}) and (\ref{g22}) by $\,2\:\!\cf$ and
$\,\cf \beta_0 - 14\,\cfs$, respectively. The $\zeta$-function contributions, 
in particular, are the same. For comparison and use with future four-loop
computations, we also carry out the inverse Mellin transform of these results 
for $\,n=4$. This leads to 
\bea
  c^{\,(4)}_{L,q}(x) &\!\! =\! & \mbox{Eqn.~(\ref{cL4x})} 
\nn \\[0.5mm] & & \mbox{\hspn\hspn}
  \: - \: (1-x) \Bigg( \,
    {16 \over 3}\: \cff\: L_x^{\,6}
  \: + \: \Bigg\{
    [\, 8 - 64\, \z2 ] \,\cff
    - \Bigg[ {728 \over 9} - 32\, \z2 \Bigg] \: \cft \ca
    + {80 \over 9}\, \cft \nf
  \Bigg\} \: L_x^{\,5}
\nn \\[1mm] & & \mbox{\hspn}
    - \: \Bigg\{ [\, 568 - 608\, \z2 + 160\, \z3 ] \,\cff
    + \Bigg[ {4544 \over 9} - {736 \over 9}\: \z2 + 208\, \z3 \Bigg]\: \cft \ca
\nn \\[1.5mm] & & \mbox{} \;
    + \Bigg[ {3388 \over 9} - {1360 \over 9}\: \z2 - 64\,\z3 \Bigg]\: \cfs \cas
    - \Bigg[ {368 \over 9} + {704 \over 9}\: \z2 \Bigg]\: \cft \nf
\nn \\[1.5mm] & & \mbox{} \; 
    - \Bigg[ {880 \over 9} - {352 \over 9}\: \z2 \Bigg]\: \cfs \ca \nf
    + {16 \over 3}\: \cfs \nfs
  \Bigg\} \: L_x^{\,4} \; + \; O(L_x^{\,3}) \Bigg) 
    \; + \; O\left( (1-x)^2 \right) \;\; .
\eea

\begin{figure}[bht]
\vspace*{-2mm}
\centerline{\epsfig{file=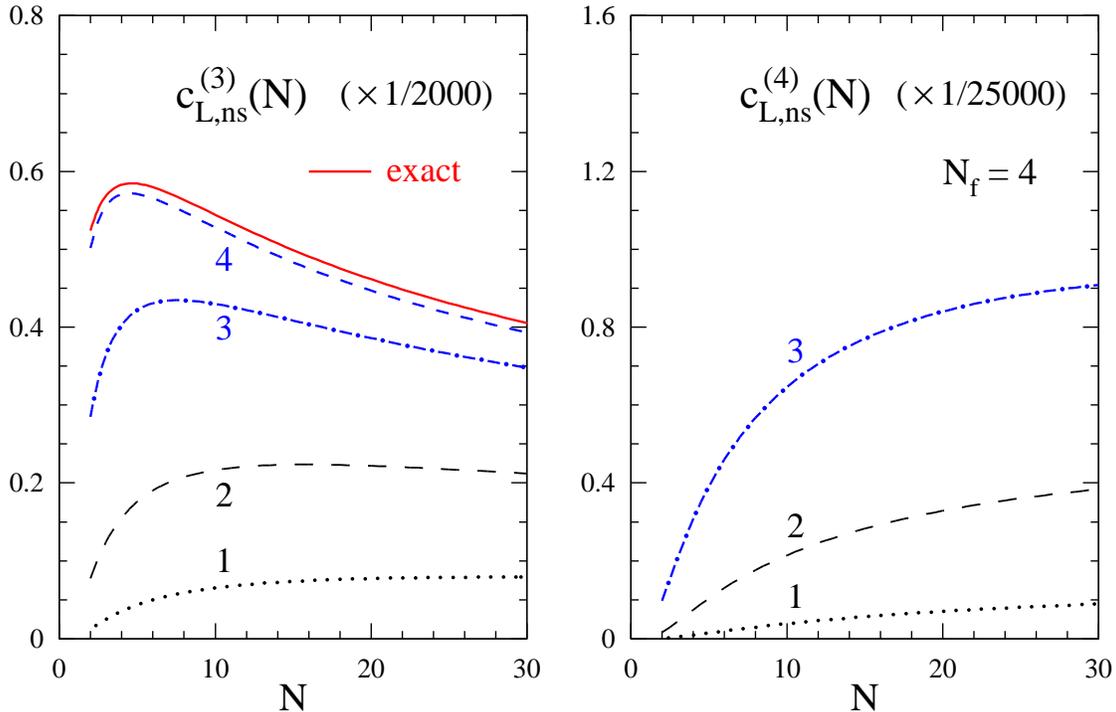,width=15.2cm,angle=0}}
\vspace{-2mm}
\caption{ \label{fig1}
 Successive large-$N$ approximations by the leading 1, 2, 3 and (left) 4 
 large-$N$ logarithms $\ln^{\,k}\widetilde{N} \equiv (\ln N + \GE)^k$ for the
 third- and fourth-order quark coefficient function of $\FL$ for four 
 flavours. Also shown is the complete third-order all-$N$ result  
 computed in Refs~\cite{Moch:2004xu,Vermaseren:2005qc}. The curves have been
 scaled to correspond to the expansion parameter $\as$ instead of
 $\ar = \as/(4\pi)$ used in our formulae.
 }
\vspace{-2mm}
\end{figure}

Finally we briefly illustrate the approximations of the $N$-space coefficient 
functions $c^{\,(n)}_{\,L,\rm ns}(N)$ in terms of the leading $N^{\,-1} 
\ln^{\,k}\widetilde{N}$ contributions (obtained from Eqs.~(\ref{cL-LL}) --
(\ref{cL-NLL}) by nullifying the Euler-Mascheroni constant $\GE$ in Eqs.~%
(\ref{cL21}) -- (\ref{cL32}), recall $\ln \,\widetilde{N} = \ln\,N + \GE$).
In the left part of Fig.~\ref{fig1} we compare the
successive approximations obtained by including one (only the $\ln^{\,4}
\widetilde{N}$ term, the curve labeled `1' in the figure), two (that and the 
$\ln^{\,3}\widetilde{N}$ term, curve 2) etc large-$N$ logarithms to the 
complete result of Refs.~\cite{Moch:2004xu,Vermaseren:2005qc}. 
We see that including all four logarithms leads to a good approximation down
to surprisingly low values of $N$, and that the highest three logarithms 
provide a reasonable first estimate at large $N$.

Our new predictions (\ref{tower1}) -- (\ref{tower3}) for the three highest
logarithms at order $\as^{\,4}$ are shown in the same manner in the right part
of Fig.~\ref{fig1}. Comparing the shape and relative size of these terms with
those of the three-loop contributions, one has to conclude that three leading
logarithms alone are insufficient for a quantitative prediction of the unknown 
coefficient function for~$\FL$. One may expect that the complete coefficient 
function exceeds the three-logarithm result in Fig.~\ref{fig1} by a factor of 
about 1.5 to 3 at $N \,\simeq\, 15 \ldots 30$. This is consistent with the
fourth-order Pad\'e predictions, e.g., 
\beq
  C_{L,\rm ns}(N=20) \:\: =\:\: 
  0.0202\,\as \:+\: 0.108\,\as^{\,2} \:+\: 0.465\,\as^{\,3} 
  \:+\: 2.0_{\,[1/1]\:\rm Pad\acute{e}\,}\,\as^{\,4} \:+\: \ldots  \:\: .
\eeq
Hence the present results are compatible with (but of course not conclusive
of) a fourth-order continuation of the very slow large-$N$ convergence of 
$\FL$ already discussed at order $\as^{\,3}$ in Ref.~\cite{Vermaseren:2005qc}.
  
To summarize, we have derived an explicit all-order resummation of the leading
and sub-leading large Mellin-$N$ contributions, $\as^{\,n}\,N^{-l} \ln^{\,k} N$
for $l = 1$ and $l = 2$, to the quark coefficient function for the longitudinal
structure function $\FL$ in deep-inelastic scattering. The resummation is
performed `bottom-up' by exploiting the absence (established to $n=3$ by the
complete results of Refs.~\cite{Moch:2004xu,Vermaseren:2005qc}, and conjectured
for all higher orders) of double-logarithmic contributions to the physical 
evolution kernel for the flavour non-singlet part of $\FL$. Specifically we 
obtain the three highest logarithms at each order $n \geq 4$, i.e., the terms 
$\as^{\,n}\: (1-x)^{l-1} \ln^{\,2n-1-k} (1-x)$ for $l\,=\, 1,\: 2$ and 
$k\,=\, 1,\: 2,\: 3$ after transformation to Bjorken-$x$ space. These 
contributions alone are not relevant for phenomenology, but will become useful
in conjunction with future higher-order calculations of, e.g., some integer-$N$
moments of this coefficient function.

With three terms per order, our present resummation has the predictive power of
a next-to-leading logarithmic exponentiation, cf.~Ref.~\cite{Moch:2005ba}. 
However, writing the results in a manner analogous to the well-known 
exponentiation of the $\as^{\,n} \ln^{\,k} N$ contributions to, e.g., the 
structure function $\Ftwo$, we notice a peculiar behaviour of the 
next-to-leading function $g_2^{}(\as \ln N)$ in the exponent: the second 
Taylor-coefficient is not, as for $F_2$, a simple function of the first and
the $\as^{\,2}$ cusp anomalous dimension, and hence the third coefficient
cannot be predicted at this point (if at all). If that coefficient could be derived in
a `top-down' approach complementary to that of this letter, then a forth 
tower of logarithms would be calculable via matching to the known (but 
presently unused) $\as^{\,3} \ln\, N$ coefficient. It might even become 
possible to achieve a full next-to-next-to-leading logarithmic accuracy which 
would provide a realistic estimate of the fourth-order large-$x$ coefficient 
function.

\noindent{\bf Acknowledgments:}
Some of our symbolic manipulations have been carried out in {\sc Form}
\cite{Vermaseren:2000nd}.
S.M. acknowledges support by the Helmholtz Gemeinschaft under contract 
VH-NG-105 and in part by the Deutsche Forschungsgemeinschaft (DFG) in 
Sonderforschungs\-be\-reich$\,$/$\,$Transregio~9. 
The \mbox{research} of A.V. has been supported by the UK Science \& Technology 
Facilities Council (STFC) under grant numbers PP/E007414/1 and ST/G00062X/1.
 
{\footnotesize
\setlength{\baselineskip}{0.5cm}

}

\end{document}